\documentclass{aa}
\usepackage{epsf,psfig,epsfig,graphics}
\begin{document}
   \title{GMRT observations of the field of INTEGRAL X-ray sources- I}
   \author{M. Pandey\inst{1}, R. K. Manchanda\inst{2}, A. P. Rao\inst{3}, P. Durouchoux\inst{4}, Ishwara-Chandra\inst{3}}

\offprints{M. Pandey : mamta@ncra.tifr.res.in}
\institute{Department of Physics, Mumbai University, Mumbai - 400 098\and
Department of Astronomy and Astrophysics, TIFR, Colaba, Mumbai -400005\and
NCRA, TIFR, Post Bag 3, Ganeshkhind, Pune - 411 007\and
CNRS FRE 2591/CEA Saclay, DSM/DAPNIA/SAP, F-91191 Gif sur Yvette Cedex, France}
 \date{Received ; accepted}
\authorrunning{M. Pandey et al.}
\titlerunning{GMRT Observations of the field of INTEGRAL X-ray sources- I}

\abstract{
The INTEGRAL observatory has discovered a variety of hard X-ray sources in the 
Galactic plane since its launch. Using GMRT, we have made repeated observations 
of these sources to search for the radio counterparts of seventeen of them at low 
frequencies. The source positions were taken from the various ATEL and IAUC
reporting their discovery. Possible radio counterparts for seven of these sources 
namely, IGR J06074$+$2205, IGR J15479$-$4529, IGR J16479$-$4514, IGR J17091$-$3624, 
IGR J18027$-$1455, IGR J18539$+$0727 and IGR J21247$+$5058 were detected within 
3$\sigma$ of the  position uncertainty derived from the INTEGRAL observations. The 
offset in the radio position was calculated using the positions mentioned in the ATEL. 
We have also analyzed the available NVSS images for a few of these fields at 1.4 GHz 
along with our observations. In this paper we present the  radio images and the best 
fit positions for the positive detections. The  X-ray variability for few of the 
sources within the time scales of 100 s to 1 ks as seen in the RXTE/PCA light curves 
suggests their Galactic origin and possible binary nature. We discuss briefly the 
characteristics of these sources from the available information at different wave bands.

}

\maketitle
\section{Introduction}
Since its launch in Oct 2002 the International Gamma Ray Astrophysics Laboratory,
({\emph{INTEGRAL}}), has discovered a number of new hard X-ray/ soft $\gamma$-ray 
sources\footnote{See http://isdc.unige.ch/$\sim$rodrigue/html/igrsources.html for an 
updated information on the list of all {\emph{INTEGRAL}} sources} within the Galactic plane 
mainly towards the Galactic center region. These sources were discovered with the IBIS 
instrument, which has a large field of view of $29^\circ\times29^\circ$ and a sensitivity 
up to 1 MeV (Lebrun {et al.} 2003). Most of these new Galactic sources are soft gamma-ray 
emitters in the 20 $--$ 100 keV energy range and have been identified as X-ray binary 
systems (XRBs) (Bird {et al.} 2004).
\par
The Galactic XRBs sources are accreting binary systems with black hole (BH) or neutron 
star (NS) as the compact object. Apart from persistent X-ray emission from a number of 
these binaries, a large fraction of these systems are transient in nature,  which brighten 
up on occasion, sporadically and each source has some unique characteristics. Since 
observations of superluminal outflows during the transient burst of GRS 1915$+$105
(Mirabel \& Rodr\'{\i}guez, 1994), a new subclass, characterized by the presence of jets
in X-ray binaries, has been established. The radio observations of some of these
Galactic sources during quiescent and flaring modes using radio interferometers suggest 
their morphology is similar to quasars, hence they are named as microquasars. A compact 
core and two sided radio jets have been commonly observed in these sources during VLBI 
observations(Mirabel {et al.} 1999). In most of these sources a compact object is believed 
to be a black hole candidate (BHC); however, for Sco X-1, CI Cam and Cir X-1, the compact
companion is determined to be a neutron star. Interesting variability pattern in the radio 
light curve were seen for  GRS 1915$+$105 (Mirabel {et al.} 1994, 1997, 1998). The X-ray 
emission from this source has also revealed a rich class of flux variability and temporal 
features (Belloni {et al.} 1997).

\begin{table*}
\begin{center}
\caption{\bf List of target {\emph{INTEGRAL}} sources observed with GMRT}
\begin{tabular}{llllllll}
\hline
\hline
Source           &Type        &Integral     &Variable             &X-ray     &X-ray    &GMRT         &\\
                 &            &Pos. Unc     &100 s-1 ks             &Flux      &State    &detection    &\\
                 &            &1.6$\sigma$  &IBIS/ISGRI           &(mCrab)   &during   &within X-ray &\\
                 &            &             &                     &15 $--$ 40 keV &GMRT obs.&3$\sigma$error box&\\
\hline
IGR J06074$+$2205  &-            &2$'$           &Yes                 &15    &high soft&Yes&\\
IGR J15479$-$4529  &-            &2$'$           &Yes                 &2     &-        &Yes&\\
IGR J16316$-$4028  &-            &3$'$           &Yes                 &25    &-        &Yes&\\
IGR J16318$-$4848  &sgB(e)(HMXB) &4$'$$'$        &Yes                 &50-100&high soft&No &\\
IGR J16320$-$4751  &HMXB(?)      &4$'$$'$        &Yes                 &10-50 &-        &No &\\
IGR J16418$-$4532  &-            &2$'$           &-                   &2     &-        &No &\\
IGR J16479$-$4514  &             &3$'$           &Yes                 &12    &high soft&Yes&\\
IGR J17091$-$3624  &XB$^{1}$(BHC)&0.8$'$         &Yes                 &40    &low hard &Yes&\\
                   &?            &               &                    &-     &-        &-  &\\
IGR J17391$-$3021  &Be/(NS)HMXB$^{2}$&1$'$$'$    &Yes                 &150   &-        &No &\\
                   &?            &Chandra        &                    &-     &         &   &\\
IGR J17544$-$2619  &HMXB(?)      &4$'$$'$        &Yes                 &60    &high soft&No &\\
IGR J17597$-$2201  &Neutron Star, LMXB&2$'$      &Yes                 &5     &-        &No &\\
IGR J18027$-$1455  &-            &2$'$           &Yes                 &-     &high soft&Yes&\\
IGR J18325$-$0756  &-            &2-3$'$         &Yes                 &-     &-        &No &\\
IGR J18483$-$0311  &-            &2$'$           &Yes                 &10    &-        &No &\\
IGR J18539$+$0727  &XB$^{3}$     &3$'$           &Yes                 &20    &low hard &Yes&\\
                   &(BHC)        &               &                    &-     &         &   &\\
IGR J19140$+$0951  &LMXB$^{4}$   &1.3$'$         &Yes                 &50$--$100&high soft&No&\\
                   &(BHC)        &               &                    &-     &         &   &\\
IGR J21247$+$5058  &Radio Galaxy &2$'$           &Yes                 &-     &low hard &Yes&\\
\hline
\end{tabular}
\footnotemark{M. Revnivtsev {et al.} 2003a$^1$, Smith {et al.} 2003$^2$, M. Revnivtsev {et al.} 2003b$^3$, 
Schultz {et al.} 2004$^4$}
\end{center}
\end{table*}

\par
We have been carrying out a programme of monitoring of the new {\emph{INTEGRAL}} sources
since Aug 2003 with the Giant Metrewave Radio Telescope, (GMRT) with a view:
\begin{enumerate}
\item to map the field of new X-ray sources at low frequencies in order to find the radio
counterparts and provide the exact position coordinates with the arcsec resolution of
GMRT at meter wavelength.
\item to derive a possible association of the source by combining the radio, infrared and
X-ray data and to establish the nature (low mass X-ray binary (LMXB) or high mass X-ray binary (HMXB)) 
of the source based on the magnitude of any infrared counterpart. Both these systems have NS or BH as 
one of the component. In the case of HMXB the other component is a massive star, usually a Be 
star or a blue super-giant of $\sim$ 8 $--$ 20 M$\sun$. And for LMXBs the second component (donor) 
is a main sequence star. In some cases the donor can either be a degenerate or a evolved 
(sub-giant or red giant).
\item to distinguish the Galactic (compact) and extragalactic (extended) sources based on their
statistically known radio morphology.
\end{enumerate}
The radio data at different frequencies can help in unveiling the nature of these new sources
and physics underlying the emission/absorption processes (Revnivtsev {et al.} 2003a).
\par
In this paper we report the results obtained from radio observations performed on seventeen
new sources with GMRT at 0.61 GHz and 1.28 GHz. The selection criteria applied for possible
radio-X-ray association discussed below is derived from our rigorous study of the known
XRBs and BH jet sources. This was initiated in 2001 using GMRT (Pandey {et al.} 2004)
with a view to establish the long term spectral behavior of these sources at radio frequencies in
correlation with their X-ray properties.
\par
The selection criteria for a genuine and possible radio-X-ray association of the X-ray sources
is as follows:
\begin{enumerate}
\item  The number of extragalactic sources with flux densities above 2.5 mJy in the sky at
0.61 GHz in 10$'$ $\times$ 10$'$ area is typically $\sim$ 5 whereas the number of sources expected above the flux
density of 1.1 mJy in 10$'$ $\times$ 10$'$ area at 1.28 GHz is typically $\sim$ 2 (Condon {et al.} 1999).
The number of sources decrease with: (i) the increase in flux density and (ii) 
decrease in the field area. The above source counts (number of sources) are used as typical limit for 
the extragalactic sources in the given area. The number of sources above this limit in the given area 
are considered as Galactic sources. The occurance of number of Galactic radio sources in
the field of {\emph{INTEGRAL}} source within the X-ray position error circle is statistically higher in number as 
compared to extragalactic radio sources.
Also this source counts varies from the Galactic center to further away along the Galactic plane
with lower probability in the latter region. 
\item The X-ray binaries have compact radio morphology with the radio emission being 
non-thermal in origin. The variability in the radio flux density is a prominent feature 
in these sources.  The radio spectra of microquasars show flattening or power law decay at high radio 
frequencies with radio emission coming from the jet, e.g. GRS 1915$+$105, SS433, 
Cyg X-1, and spectral turnover at low radio frequencies with radio emissions coming from 
the radio lobe (Ishwara-Chandra {et al.} 2004). The radio emission is quenched when the 
source is in high soft X-ray state.
Also the radio emission from few of these sources are synchrotron self absorbed below
1.28 GHz, e.g. X1812$-$121, XTE J1720$-$318, 1E1740.7$-$294 (Pandey {et al.} 2004). The X-ray 
characteristics include prominent fluorescent emission from the iron line at 6.4 keV and a
hard X-ray tail showing rapid variability of the order of $\sim$ 10 $--$ 100 s. 
The infrared and optical counterparts are associated with the X-ray binary source. The infrared
emission originates from circumstellar disks undergoing major expansion coincident with phases of X-ray
activity and the evaporation of the disk material due to heating gives rise to optical emission. 
Thus a compact radio source within the 3$\sigma$ X-ray position uncertainty limit satisfying 
the above criteria is very likely to be associated with the Galactic X-ray source.
\item The radio sources with extended radio morphology and with their radio emission 
being non-thermal in origin have spectral index, $\alpha \sim -0.7$ 
(S$\sb{\nu}$=$\nu^{\alpha}$), are classified as extragalactic 
radio sources. Their radio emission is mostly non 
variable. The X-ray light curve shows variability on larger timescales;  however,
variability of the order of $\sim$ 100 s is reported for few extragalactic sources. Thus a radio 
source within the 3$\sigma$ X-ray position uncertainty limit satisfying the above criteria 
is most likely not associated with the Galactic X-ray source.
\end{enumerate}

\par
The observations on these seventeen sources presented here were made in the Aug 2003 observation cycle 
with GMRT. Further observations of 23 sources belonging to this class were conducted in 
Aug 2004 and the data analysis is in progress. The target sources are those discovered 
during the {\emph{INTEGRAL}} observation scans on the Galactic plane, and during guest 
observations and Crab calibrations; they are listed in Table 1. 

\section{Observations and Analysis}
The radio observations were carried out at 0.61 GHz and 1.28 GHz with a bandwidth of 16/32 MHz 
using the GMRT (Swarup  1991). 
The Half power beam width for the GMRT 
antenna at 0.61 GHz is $\sim$ 43$'$ and at 1.28 GHz is $\sim$ 26$'$. The measured 
position offset of the possible GMRT radio counterparts with respect to the X-ray sources 
is below $\sim$ 5$'$ , hence primary beam corrections do not contribute a significant change 
to the images. The gain decreases significantly in the field at low radio frequencies, i.e 
0.61 GHz, hence gain depression corrections or system temperature corrections were applied 
to the flux densities of the radio counterparts. The flux density scale was set by observing 
the primary calibrators 3C286, 3C147 and 3C48. Phase calibrators were observed near the target 
source for $\sim$ 5 min scans interleaved with 25 min scans on the {\emph{INTEGRAL}} sources. 
The sample time was 16 s. The data recorded from GMRT were converted into FITS files and 
analyzed using the Astronomical Image Processing System (AIPS). A self calibration on the 
data was performed to correct for phase related errors and improve the image quality. 7 $--$ 10\% 
of the data was affected by interference and was carefully removed. Ionospheric scintillation effects 
were not seen on the 0.61 and 1.28 GHz data during our observations.
Possible radio counterparts were detected for seven out of seventeen sources and for other sources no 
possible radio counterpart was detected within the 3$\sigma$ error box of the X-ray source. 
Table 2 summarizes our results for the new sources along with best-fit radio positions for the 
counterpart and the position offsets of this counterpart with respect to the X-ray positions. 
Column 4 gives the peak and total flux density for the point and extended source respectively. 
The rms noise given in column 5 of Table 2 corresponds to the average background  noise in the 
image field  and  is higher in the Galactic plane. In the radio images presented below, 
the boxes show the known field sources (X-ray, gamma-ray, optical, infrared, radio) from the 
files available from the NASA extragalactic Database (NED), 2MASS and Digital Sky Survey (DSS). 
The number of infrared 
and optical sources expected within a 10$'$ area in the Galactic region was derived from the 
image obtained from FITS files available from the 2MASS and DSS surveys at: 
http://skys.gsfc.nasa.gov/cgi-bin/skvadvanced.pl. The expected number of infrared and optical 
sources are $\sim$ 6 and $\sim$ 1.2 respectively; however, in the radio images shown in this 
paper, only the sources in close proximity to the radio source have been marked. We also 
discuss briefly the spectral properties in the radio band (non-simultaneous radio observations)
of the radio counterparts on which other data are available. The bold circle marked with `A' 
only indicate a possible GMRT radio counterpart close to the X-ray source. We have grouped 
sources with a genuine radio-X-ray associations, no associations and no radio information, 
based on our results in Sec. 3.
 
\section{Results}
Imaging at radio wavelengths are specially important since interferometry has the intrinsic 
capability to provide a precise sub-arcsec position measurements, thus improving the error 
boxes obtained from X-ray data. In addition, interstellar extinction has very little 
effect at low radio wavelengths. 

\subsection{X-ray sources with a radio source in the field and radio-X-ray association:}

\noindent{\bf IGR J17091$-$3624 :}
This microquasar was discovered at a flux density level of $\sim$ 20 mCrab between 40 and 
100 keV (Kuulkers {et al.} 2003). BeppoSAX  detected the source at a flux density level 
of 14 $--$ 20 mCrab in the 2 $--$ 10 keV band, suggesting a very steep photon spectrum 
consistent with a spectral index of 3.0$\pm$0.4 or kT = 4.3$\pm$1.4 keV from the thermal 
bremsstrahlung (Zand {et al.} 2003). During a 3 ks observation of the source with RXTE/PCA 
in the 2 $--$ 20 keV band, the observed flux density was $\sim$ 4 mCrab. The X-ray light 
curve of the source shows a large variability on time scales from several tenths  
to several tens of seconds. The energy spectrum of the source fitted a power law index of 
1.43$\pm$ 0.03. No visible absorption was detected at the lower energies (Lutovinov \& 
Revnivtsev 2003). The combined X-ray data from BeppoSAX and {\emph{INTEGRAL}} also suggest 
a variable nature of the source with flare episodes in Apr 2003 and Sep 2003 
(Kuulkers {et al.} 2003). The radio emission from the source region was detected during 
observations with the Very Large Array (VLA) at 4.9 GHz with a flux density of 1.8$\pm$0.3 
mJy and they showed significant variability in flux density (Rupen {et al.} 2003a).
\begin{figure}[h]
\begin{center}
\psfig{figure=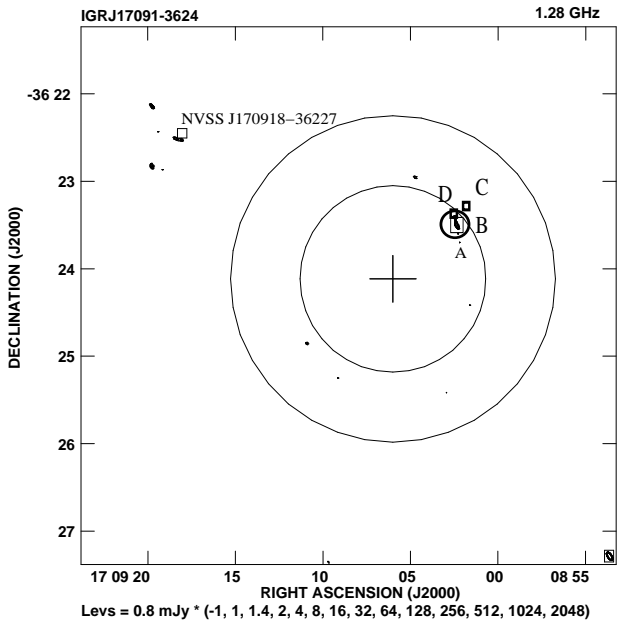,width=8.5cm,height=8.5cm,angle=0,clip=}
\end{center}
\caption{\footnotesize
{GMRT image of IGR J17091$-$3624 at 1.28 GHz with the {\emph{INTEGRAL}} position marked
 with $+$. The small and large circles shows the {\emph{INTEGRAL}} uncertainty error circles
of 1.6$\sigma$ and 3$\sigma$. The bold circle marked with `A' show the radio source detected
by GMRT. The boxes show the known field sources, B: NVSS 17092$-$3624, C: DSS J170901$-$3623 
and D: 2MASS J170902$-$3623}}
\label{fig.1.}
\end{figure}

In our observation with GMRT, a radio source was detected within the 1.6$\sigma$ position 
uncertainty of the X-ray sources at a radio flux density of 1.10$\pm$0.16 mJy at 1.28 GHz, 
at a J2000 position of RA: 17h 09m 02.3s and DEC: $-$36d 23$'$ 33$'$$'$. The rms noise was 
0.19 mJy b$^{-1}$. The flux density at 0.61 GHz was measured to be 2.50$\pm$0.73 mJy.

\begin{figure}[h]
\begin{center}
\psfig{figure=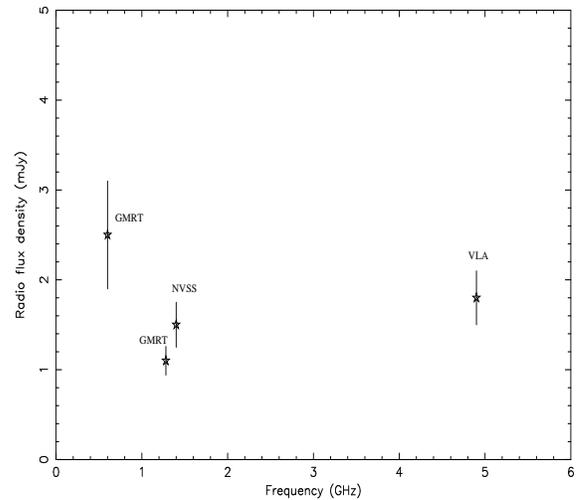,width=8.5cm,height=9.5cm,angle=-90}
\end{center}
\caption{\footnotesize {Radio spectrum for IGR J17091$-$3624.}}
\label{fig.1.}
\end{figure}

The GMRT image of the field is shown in Fig. 1. The source morphology is consistent with a point 
source. The compact nature of the radio source is also inferred from the gaussian fit to the 
source flux density using the AIPS routine `JMFIT'. In our analysis of the  NVSS FITS 
files,  we have detected a weak radio source with a flux density of $\sim$ 1.50$\pm$0.25 mJy 
at the same position. The radio source marked `A' in the figure lies within the 
{\emph{INTEGRAL}} position error circle of the source and is the most likely radio counterpart 
of the X-ray source. The GMRT (A) and  NVSS (B) radio counterparts are coincident in position. 
The infrared source, (D) and the optical source, (C) coincide in 
position with the radio source. Thus we confirm that the source (B), (C) and (D) are associated 
with the X-ray binary source. It is reasonble to think that the 
infrared emission comes from a circumstellar disk undergoing major expansion coincident with 
phases of X-ray activity and the optical emissions are due to heating and evaporation of the 
disk material. In order to enhance the significance of the possible association of the GMRT 
source with the X-ray object, we have plotted the radio flux density measurements of the source 
region at GMRT frequencies of 0.61 and 1.28 GHz in Fig. 2. The observed flux density values for 
the NVSS source at 1.4 GHz and VLA flux density value at 4.9 GHz are also plotted in the figure 
for reference. However, it should be noted that the observations are non-simultaneous. 
The spectral fit to the GMRT data corresponds to the falling power law of the form ({\it S 
$\propto \nu^{\alpha}$}) with $\alpha$ = $-$0.62. This is consistent with the radio emission from
the other X-ray binary sources. It should be noted that if the high frequency observations are 
associated with the same emission region, the radio spectral fit will require a clear turn 
around to positive slope with $\alpha = +2.73$. This suggests a composite of two separate radio 
emitting regions for the source.
\\\\
\noindent{\bf  IGR J18027$-$1455 :}                                                    
This source is an AGN candidate located in the Norma arm region (Walter {et al.} 2004, 
Masetti {et al.} 2004). A faint ROSAT X-ray source 1RXS J180245.5$-$145432, lies 
well within the IBIS/ISGRI error 
circle. Two weak radio sources were discovered in the NVSS field at 1.4 GHz (Combi {et al.} 2004, 
Ribo {et al.} 2003). The first source, NVSS J180239$-$145453, located at
J2000.0 ICRS position of RA: 18h 02m 39.94s($\pm$0.25s), DEC: $-$14d 54$'$ 53.6$'$$'$($\pm$3.3$'$$'$) and 0.51
 arcmin away from the X-ray source has a flux density of 6.9 mJy. This object has an infrared 
counterpart 2MASS J180239$-$145453. The second source, NVSS J180247$-$145451 at J2000.0 ICRS position of 
RA: 18h 02m 47.37s($\pm$0.13s), DEC: $-$14d 54$'$ 51.6$'$$'$($\pm$2.2$'$$'$) is located 0.28 arcmin away from the 
X-ray position and has a flux density of 10.5 mJy. The field is further complicated by the presence 
of an extended IR source, 2MASXi J1802473$-$145454, with coordinates RA: 18h 02m 47.3s, DEC: $-$14d 54$'$ 55$'$$'$ 
and magnitudes $J=13.18$, $H=12.02$, $K=10.94$, within the 2$\sigma$ position error ellipse of the second 
radio source. The average optical magnitudes of this IR source are  $B=19.3$, $R=14.9$ and $I=13.8$ 
in the USNO-B1.0 catalog. The source has an optical counterpart DSS J180241$-$145453.  
The sources NVSS J180247$-$145451 and NVSS J180239$-$145453.6, 
2MASXi J1802473$-$145454, DSS J180241$-$145453 and 1RXS J180245.5$-$145432 are therefore, the equally 
likely counterparts of IGR J18027$-$1455.
\begin{figure}[h]                                                                    
\begin{center}
\psfig{figure=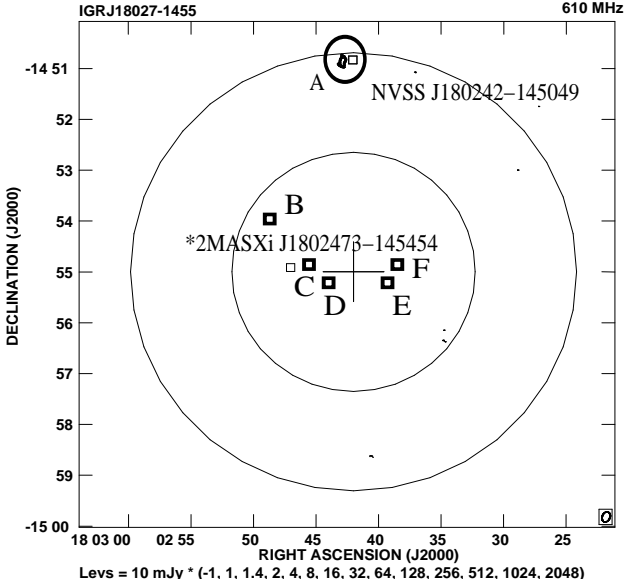,width=8.5cm,height=8.0cm,angle=0, clip=}                             
\end{center}
\caption{\footnotesize { GMRT image of IGR J18027$-$1455 at 0.61 GHz with the {\emph{INTEGRAL}} 
position marked with $+$. The small and large circles shows the {\emph{INTEGRAL}} uncertainty 
error circles of 1.6$\sigma$ and 3$\sigma$. The bold circle marked with `A' shows a radio 
source detected by GMRT. The boxes show known field sources, B: 1RXS J180245.5$-$145432, 
C: NVSS J180247$-$145451, D: DSS J180241$-$145453, E: NVSS J180239$-$145453.6 and F: 2MASS J180239$-$145453.}}
\label{fig.1.}
\end{figure}
The GMRT observations were made at 0.61 GHz and image of the field is shown in Fig. 3. It is seen
from the figure  that GMRT data did not yield any significant flux from the directions of
NVSS J180239$-$145453 (E), 2MASS source (F) or ROSAT (B) source position. The only point radio source 
seen (A), in the GMRT data was at a flux density of  10.72$\pm$2.25 mJy and is coincident with the NVSS source 
J180242$-$145049. The GMRT position for the radio source is RA: 18h 02m 42.75s and 
DEC: $-$14d 50$'$ 49.21$'$$'$ with position offset 4.18$'$ with respect to the {\emph{INTEGRAL}} 
position. The rms noise was 2.3 mJy b$^{-1}$, and the NVSS flux density for the source is 12.3 mJy. 
With the detection of higher flux density at 0.61 GHz compared with NVSS data, we can infer that NVSS 
J180242$-$145049 is a flat spectrum source in the 0.61 GHz to 1.4 GHz range, assuming the source
is persistent in nature.
  
Non detection of the 2 radio sources closest to the {\emph{INTEGRAL}} X-ray source position during 
our observations, suggests that the radio sources show spectral turnover at low frequencies or are 
variable in nature. However in both the cases, a non-thermal origin of the radio emission is 
clearly favoured.
Thus considering the NVSS flux density at 1.4 GHz for the source 
NVSS J180247$-$145451 and the upper limit of $<$7 mJy from GMRT observations 
at 0.6 GHz, the spectral index derived is, $\alpha$ = $-$0.46, typical for binary jet sources.
The RXTE/ASM light curve for the hardness ratio of the X-ray source suggests that it was in a 
high soft state during our observations. 
Thus, NVSS J180247$-$145451 and NVSS J180239$-$145453.6 are the radio counterparts of the X-ray source. 
The non-detection of NVSS J180239$-$145453 (E) and NVSS J180247$-$145451 (C) with GMRT in May 2004 
and a clear detection in the NVSS survey is consistent with the binary nature of the X-ray source.
However very recent observation of IGR J18027$-$1455 region with the GMRT (2nd Mar 2005) shows 
an evolution of the previous trends followed by the source in the radio regime. We have detected radio 
emissions at 0.61 GHz in the direction of the {\emph{INTEGRAL}} source and coincident in position 
with the NVSS (E) and (C) sources; see Fig. 3. The radio morphology reveals the presences of 
radio lobes. The (E) and (C) sources have GMRT flux densities of 9.00$\pm$0.33 and 
5.00$\pm$0.35 mJy, respectively. A clear low frequency radio variability is established for the 
source via our observations. The ASM light curve for the source shows a flaring episode 
during MJD 53350 $--$ 53360 i.e 1st $--$ 11th Dec 2004. This enhances the significance of the radio--X-ray 
association.\\\\

\noindent{\bf IGR J18539$+$0727 :}
This source was discovered with the IBIS/ISGRI detector in Apr 2003 
(Lutovinov {et al.} 2003). With the measured photon fluxes of 20 mCrab 
between 15 and 40 keV and $\sim$ 20 mCrab between 40 and 100 keV, IGR J18539$+$0727  
has the most unusual flat spectrum among the known X-ray sources in hard X-ray band. 
In a later observation of the source with RXTE/PCA, an average flux density level 
of $\sim$ 6 mCrab was found with large variability on time scales from milli-seconds to 
seconds, but without any prominent  features associated with quasi-periodic oscillations (QPO). 
The energy spectrum of the source fits a power law with spectral index $\sim$ $-$1.5, and 
appreciable low energy attenuation due to neutral hydrogen of 
$n_H\sim (1.5\pm 0.4) \times 10^{22}$ and a fluorescent iron emission line at 6.4 keV.
The observed spectral and temporal characteristics of the X-ray source fits well 
with the standard model of AGNs (Ceballos {et al.} 2004).

\begin{figure}[h]
\begin{center}
\psfig{figure=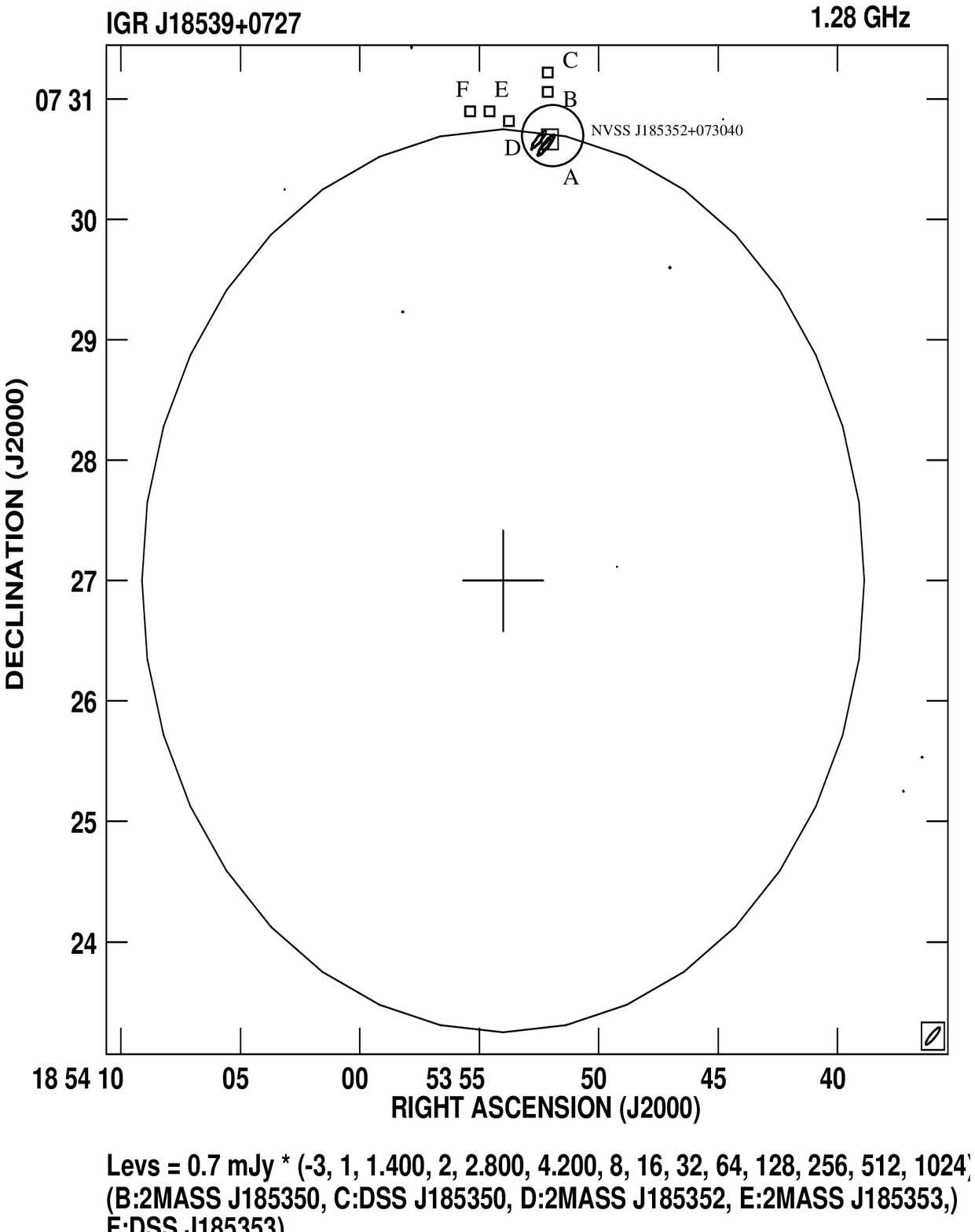,height=8.2cm,width=8.5cm,angle=0,clip=}
\psfig{figure=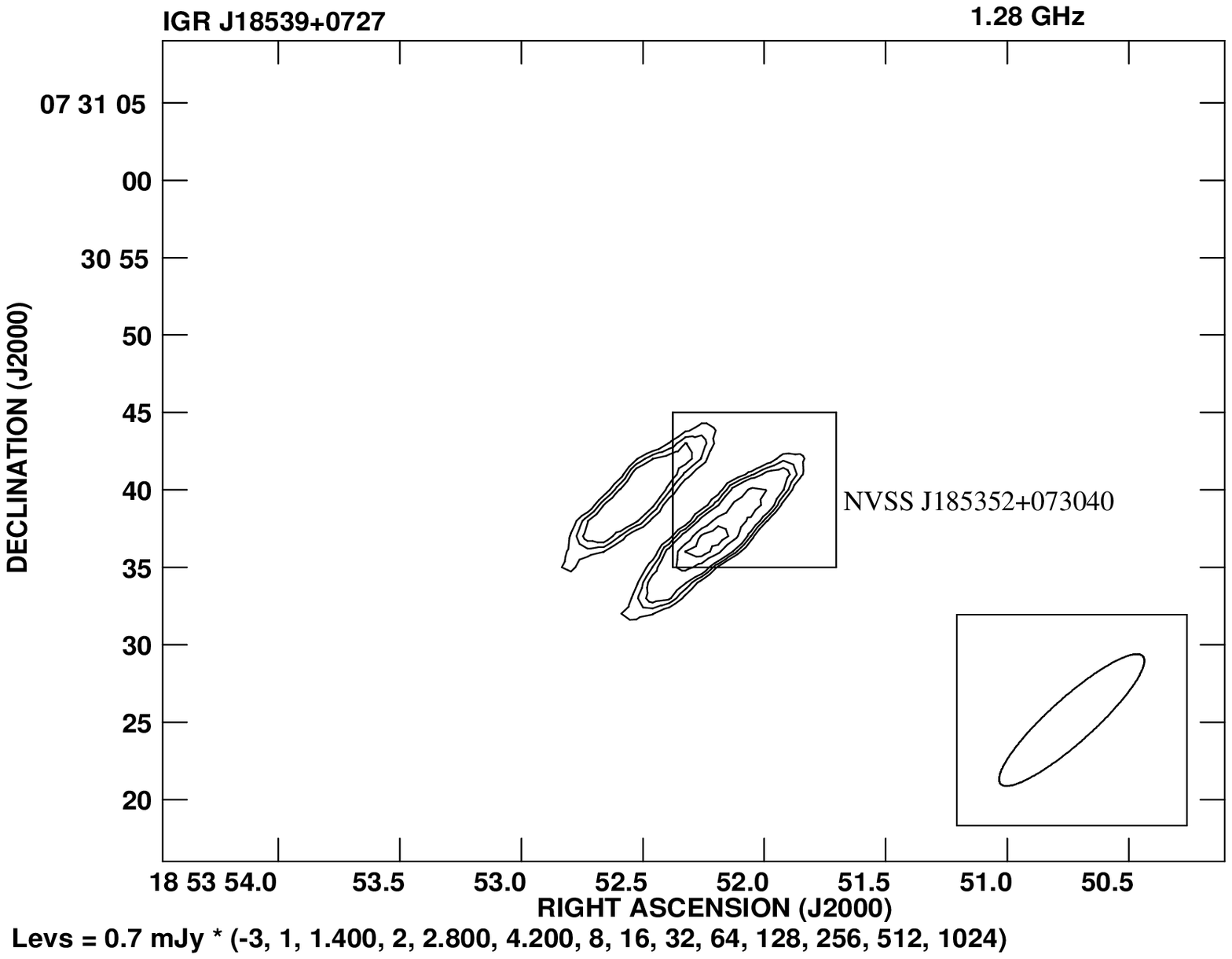,height=8.5cm,width=8.5cm,angle=0}
\end{center}
\caption{\footnotesize
{{\it Top} fig shows the GMRT image of IGR J18539$+$0727 at 1.28 GHz with the 
{\emph{INTEGRAL}} position marked with $+$. The circle shows the
{\emph{INTEGRAL}} uncertainty error circle of 1.6$\sigma$.
The bold circle marked with `A' shows the radio source detected by GMRT. 
{\it Below} fig shows the blown up image of GMRT radio counterpart.
The boxes show the known field sources, B: 2MASS J18535$+$0731, C: DSS J18535$+$07315, 
D: 2MASS J185352$+$07305, E: 2MASS J185353$+$07305 and F: DSS J185353$+$07305}}
\label{fig.1.}
\end{figure}

During GMRT observations of the field, a fully resolved double point source with 
a flux density of $\sim$ 5.25$\pm$0.25 mJy and $\sim$ 5.85$\pm$0.50 mJy was 
detected from the J2000 position of, RA: 18h 53m 52.54s$\pm$0.03 and DEC: $+$07d 30$'$ 39.07$'$$'$$\pm$0.43 
and RA: 18h 53m 52.19s$\pm$0.023 and DEC: $+$07d 30$'$ 37.31$'$$'$$\pm$0.32, respectively, which lies within
the 1.6$\sigma$ position error circle of 3$'$ of the X-ray source. The radio morphology resembles 
IGR J18027$-$1455 or an AGN. The computed position offset from the field center is 3.82$'$ 
and 3.64$'$ respectively as shown in Fig. 4. The GMRT source is coincident with the NVSS 
unresolved radio source in the field, for which the flux density is estimated as $\sim$ 6 mJy at 1.4 GHz. 
The radio flux density measurements, although done at different epochs is consistent with a flat 
spectrum source between 1.28 and 1.4 GHz and its persistent  nature. Fig. 4 also shows the 2MASS J18535$+$0731 
(infrared, E) and DSS J18535$+$07315 (optical, C) coincident in position with the radio source, thus 
suggesting a possible association with the radio source. A power law X-ray spectrum with a strong 
photoabsorption at low energies and fluorescent iron emission line at 6.4 keV clearly points to a 
binary nature of IGR J18539$+$0727, in which the emission may arise due to the reflection of the 
hard X-radiation of inner accretion-flow regions from an optically thick, cold accretion disk. 
Furthermore, the  power density spectrum of the source shows a cutoff  below  a frequency of 
$\sim$ 0.01 Hz, which places it in the compact companion to a BHC category 
(Van der Klis {et al.} 1999). The presence of infrared and optical counterparts also supports 
the disk-jet morphology. Thus the hard X-ray source is associated with the radio counterpart 
detected by GMRT and we conclude from our result that IGR J18539$+$0727 is an AGN. 

\subsection{X-ray sources with a radio source in the field and no radio-X-ray association:}

\noindent{\bf  IGR J06074$+$2205 :} 
The transient X-ray source IGR J06074$+$2205 was detected with a flux density of $\sim$ 7 mCrab 
between 3 and 10 keV by {\emph{INTEGRAL}} (Chenevez {et al.} 2004). The hardness ratio 
derived from the RXTE/ASM light curve suggests that the source was in a high state during our 
observations. The EGRET source 3EG J0617$+$2238 lies close to the X-ray source; however, it is likely 
to be associated 
with the SNR IC443 (Torres {et al.} 2003). GMRT observations for the source were 
made on 4 occasions during Feb $--$ May 2004 and a radio source was discovered within 
the {\emph{INTEGRAL}} position uncertainty limit, of 1.6$\sigma$ $(\sim 2$'$)$.

\begin{figure}[h]
\begin{center}
\hbox{
\psfig{figure=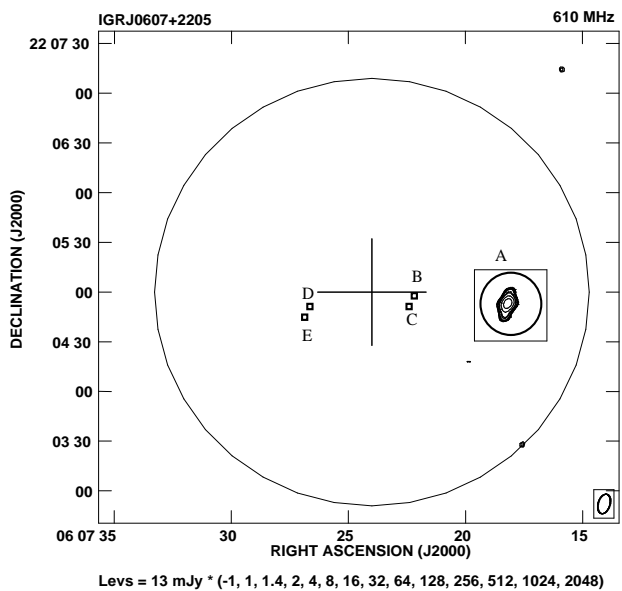,height=8cm,width=8.5cm,angle=0,clip=}
}
\end{center}
\caption{\footnotesize{GMRT image of IGR J06074$+$2205 at 0.61 GHz with the
X-ray source position in the center marked with  $+$ . The large circle represents the
{\emph{INTEGRAL}} uncertainty error circle of 1.6$\sigma$.
The bold circle marked with `A' shows the radio source detected by GMRT. The box
indicates the known NED field source, A: NVSS J060718$+$220452,  B: DSS J060722$+$220500, 
C: 2MASS J060722$+$220453, D: DSS J060727$+$220453 and E: 2MASS J060727$+$220445}}
\label{fig.1.}
\end{figure}

Fig. 5. shows the GMRT radio image of the IGR J06074$+$2205 field at 0.61 GHz with a 
source with a flux density level of $\sim$ 80 mJy close to the X-ray source position 
(offset of 1.29$'$ with respect to the {\emph{INTEGRAL}} position). The compact nature 
of the radio source is inferred from the gaussian fit to the source. 
The GMRT position coordinates are (J2000) RA: 06h 07m 18.45s 
and DEC: $+$22d 04$'$ 52.49$'$$'$. The source was also detected at 1.28 GHz by GMRT on 
1st Feb 2004 at a flux density of $\sim$ 36 mJy. An observation of this region with the 
Ryle Telescope  at 15 GHz on 29th Jan 2004, 3 days before our observations, showed 
an unresolved flux density of 4.0$\pm$0.2 mJy, at the NVSS J2000 position (Pooley 2004). 
The positive detection of the source at different epochs with GMRT at 0.61 GHz clearly 
suggests a persistent nature for the source in the radio region. The source has shown 
small variability in the radio flux density. 

\begin{figure}[h]
\begin{center}
\psfig{figure=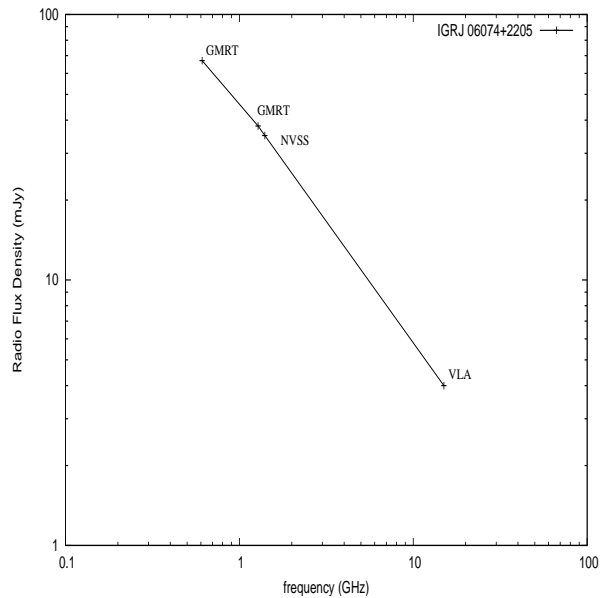,height=8cm,width=8cm,angle=0}
\end{center}
\caption{\footnotesize{Radio spectrum for IGR J06074$+$2205}}
\label{fig.1.}
\end{figure}

In the Fig. 6, we have plotted the radio spectrum  of the source using all available data.
It is seen from the figure that the data can be represented by a power-law of the form 
(S$_{\nu}\sim \nu^{\alpha}$). The spectral index at higher frequencies is derived as 
$-$0.70$\pm$0.08. A power-law nature of the radio spectrum is typical for a non thermal 
emission arising from optically thin medium similar to that produced in the jets interacting 
with the ambient medium, as seen in extragalactic sources. The hypothesis of a continuous 
jet can also lead to the X-ray emission in the hard X-ray band due to inverse comptonization 
of thermal disk photons by a corona (e.g. Markoff \& Nowak 2004). Fig. 5 also shows infrared 
(2MASS J060727$+$220453, C and 2MASS J060727$+$220445, E) and optical (DSS J060727$+$220453, D 
and DSS J060722$+$220500, B) sources present in the 1.6$\sigma$ error circle lying close to 
the X-ray source and thus showing no possible association between the radio and the 
infrared/optical sources.
\\\\
\noindent{\bf  IGR J15479$-$4529 :} 
This source was discovered by {\emph{INTEGRAL}} in Feb, 2003 
observations of Black Hole Candidates (BHC) 4U 1630$-$47 at a 
flux density level of $\sim$ 3 mCrab between 20 and 40 keV  (Tomsick {et al.} 2004). 
An X-ray source 1RXS J154814.5$-$452845,  which lies at the distance of 3$'$ from 
IGR J15479$-$4529 has been detected during  XMM-Newton observation of the source 
field in the energy range below  10 keV (Haberl {et al.} 2002). \\
During the GMRT observation, a radio source  with a total flux density of 22.2 mJy 
was detected within the position error box of 3$\sigma$ of the hard  X-ray source 
and is shown in Fig. 7. The GMRT position of the source is (J2000),  
RA: 15h 47m 44.26s and DEC: $-$45d 32$'$ 35.22$'$$'$. The rms noise in the GMRT data 
was 0.5 mJy b$^{-1}$ in the source direction. The position offset with respect to the 
{\emph{INTEGRAL}} position is 3.98$'$. The radio morphology of the source suggests 
that the source is extended in nature. No NVSS image is available for comparison 
to our measurement. In the absence of RXTE/ASM light curve, it is difficult to 
comment whether the X-ray source was in the high soft or low hard state during 
our measurements.

\begin{figure}[h]
\begin{center}
\hbox{
\psfig{figure=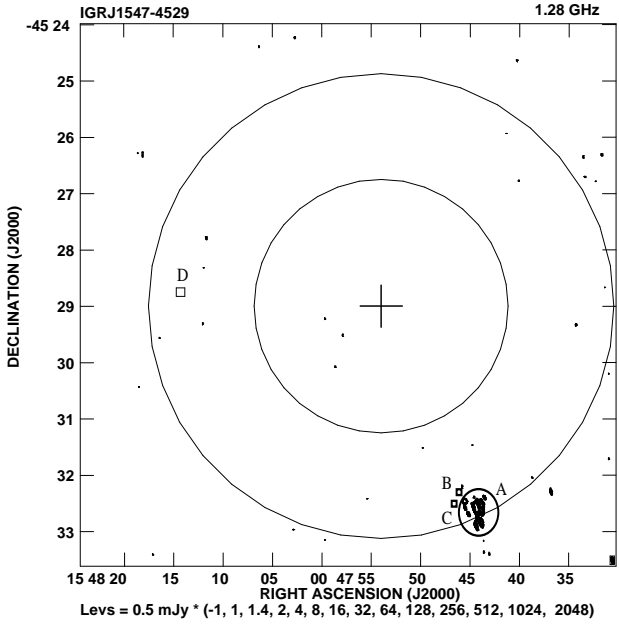,height=8.5cm,width=8.5cm,angle=0,clip=}
}
\end{center}
\caption{\footnotesize{ GMRT image of IGR J15479$-$4529 at 1.28 GHz with 
the {\emph{INTEGRAL}} position marked with  $+$ in the centre. The small and 
large circles in the figure represent the {\emph{INTEGRAL}} position 
uncertainty error circles of 1.6$\sigma$ and 3$\sigma$. The bold circle marked 
with `A' shows the radio source detected by GMRT. The boxes show the known 
NED field sources, B: DSS J1547$-$4532, C: 2MASS J1547$-$4532 and D: IRXS J154814.5$-$452845}}
\label{fig.1.}
\end{figure}

It is seen from the figure that the position coordinates of the radio source do 
not coincide with those of IRXS J154814.5$-$452845 (D) but they coincide with those of 
the infrared (2MASS J1547$-$4532, C) and optical (DSS J1547$-$4532, B) sources in position, 
thus suggesting a possible association of the radio and optical/infrared sources. From the 
above information we conclude that the radio source is not associated with the X-ray source. 
Since the X-ray counterpart is expected to fall in 90\% of the {\emph{INTEGRAL}} position 
error circle, i.e, within a distance of 1.6$\sigma$, this X-ray source may be a possible 
candidate for further observations to look for counterparts.
\\\\ 
\noindent{\bf  IGR J16479$-$4514 :} 
is a weak hard X-ray transient source with an observed flux 
density of $\sim$ 8 mCrab between 25 and 50 keV (Molkov {et al.} 2003).
The EGRET source 3EG J1655$-$4554 (Romero {et al.} 1999) is located
close to the X-ray source. The RXTE/ASM hardness
ratio suggests that the source was in a high soft state during our observations.
The ASM light curve for the source shows that the X-ray flux density between 1st Jan and
28th Feb 2003, was relatively constant though irregular large random spikes were seen.

\begin{figure}[h]
\begin{center}
\hbox{
\psfig{figure=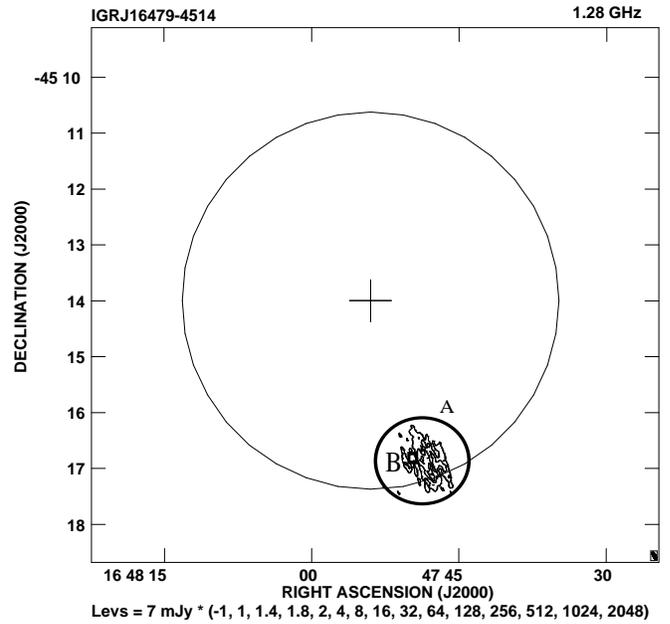,width=9cm,height=8.5cm,angle=0,clip=}
}
\end{center}
\caption{\footnotesize{GMRT image of IGR J16479$-$4514 at 1.28 GHz with the
{\emph{INTEGRAL}} position marked with  $+$ . The circle shows the {\emph{INTEGRAL}}
uncertainty error circles of 1.6$\sigma$. The bold circle marked with `A' shows the
radio source detected by GMRT and the box shows the known field source, 
B: 2MASS J164745$-$451653}}
\label{fig.1.}
\end{figure}

During our observations with GMRT in Feb 2004, a bright radio source was 
detected within the {\emph{INTEGRAL}} position uncertainty limit of 1.6$\sigma$.
The flux density recorded at 1.28 GHz is $\sim$ 2.50 Jy with the rms noise of
$\sim$ 1.24 mJy. Fig. 9. shows the 10$'$ image of the field of IGR J16479$-$4514.

The position offset is 3.30$'$ with respect to the X-ray source position. The radio
position of the source is (J2000) RA: 16h 47m 47.75s and DEC: $-$45d 17$'$ 07$'$$'$.
There is no NVSS image of the field. Radio morphology of the source suggests it is 
an HII region. While photoelectric effect heats the gas in compact HII region 
leading to ionized hydrogen, the dominant cooling
process is mainly through recombination giving rise to the free-free radio emission.
Thus radio emission from this source should be thermal in nature. The infrared 
source, 2MASS J164745$-$451653 (B) coincides in position with the radio source. Due to the extended 
nature of the radio source we conclude from our observations that the radio and infrared 
source are not associated with the Galactic X-ray source. 
\\\\
\noindent{\bf  IGR J21247+5058 :}
This source was discovered in the Norma arm region (Walter {et al.} 2004).
The X-ray source has been identified with the core of a bright radio galaxy, 4C 50.55, also
known as  GPSR 93.319$+$0.394, KR2, NRAO 659 or BG 2122$+$50 (Ribo {et al.} 2003) and is
also coincident with the IR source 2MASS 21243932$+$5058259. The optical spectrum of the
source is very peculiar, having a broad hump around 6700 \AA   of  H$\alpha$ line typical of 
elliptical or spiral bulge (Barth {et al.} 2001). While the observed Na, Ca
and Mg features are consistent with redshift $z=0$, the identification of the value of H$\alpha$ feature
leads to a redshift $z=0.020\pm0.001$. In addition, the photometric magnitudes indicate
an increased reddening from $R$ to near infrared band. It is therefore proposed that
the observed spectral features in the optical band are caused by the chance alignment
of a F-type star with the radio galaxy (Masseti {et al.}, 2004). The radio observations 
of the region with GMRT were carried out on 4 occasions between Apr $--$ May 2004.
The observation details are listed in Table. 2. The radio flux density of $\sim$ 160 
mJy at 0.61 GHz was observed on each occasion from the core (Table. 2). 

\begin{figure}[h]
\begin{center}
\psfig{figure=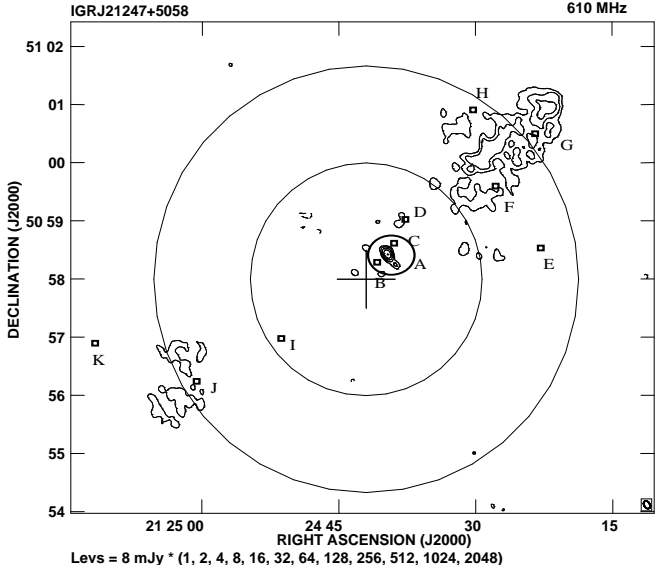,height=7.8cm,width=9cm,angle=0,clip=}
\end{center}
\caption{\footnotesize
{A high resolution GMRT image of IGR J21247$+$5058 at 0.61 GHz. The core is marked with `A'.
The X-ray position is marked with $+$. The boxes show the known field sources,
B: 2MASS J2124$+$5058, C: NVSS 21247$+$5058, D: 28P206, E: MG4 J2124$+$5058, F: *87GB 2122$+$5046,
G: 7C 2122$+$5047, H: NRAO 0659, I: *87GB 2123$+$5044, J: *87GB 2123$+$5043, K: MG4 J212511$+$5056}}
\label{fig.1.}
\end{figure}

In Fig. 9  we have plotted the high resolution GMRT image of the source along with the other radio sources 
detected in this region in previous surveys. It is seen from the figure
that morphology of the GMRT radio image clearly resembles that of a radio galaxy having 
a bright core marked `A' and extended radio lobes. However, it is seen from the figure 
that the field is very complex and contains multiple objects within the 3$\sigma$ 
position error circle of the {\emph{INTEGRAL}} source, and some of them are not seen in the 
GMRT data. Apart from the GMRT source marked `A', radio sources 28P206 (D)
and 87GB212310.9 (I) lie within the 1.6$\sigma$ position error circle of the X-ray source. 
However, the alignment of these sources towards the radio lobes clearly suggests that sources 
(D) and (I) represent regions of radio lobes which were 
brighter during the respective observations.
\par The observed GMRT flux density for the radio core varied between 173 mJy to 185 mJy at
0.61 GHz during our four observations, the peak flux densities for the radio lobes was
measured to be $\sim$ 1.8 Jy and $\sim$ 0.85 Jy. Using NVSS flux density values, for
the core and the extended radio lobes, the spectrum for the extended region fit a power 
law spectrum (S$_{\nu} \propto \nu^{\alpha}$) with a spectral index of $\alpha\sim -0.6$ 
which is consistent with the synchrotron emission from a optically thin medium. The spectral 
index for the radio core is $\alpha\sim 0.5$, which clearly 
points towards an inverted spectrum due to strong absorption in the optically thick medium at 
low frequencies. Our observation at low frequency (0.61 GHz) with GMRT has revealed the 
true spectral nature of the core of IGR J21247$+$5058. The discovery of the X-ray source in the 
hard X-ray band, the H$\alpha$ hump in the optical spectrum and its spectral nature in the 
radio and X-ray bands clearly suggest that IGR J21247$-$5058 is a partially obscured 
extragalactic X-ray source similar to NGC1068 and NGC4579.

\subsection{Sources with no radio detection:}

\noindent{\bf  IGR J16316$-$4028 :}
This source was discovered at a flux density level of $\sim$ 25 mCrab between 20
and 40 keV during the {\emph{INTEGRAL}} survey (Rodriguez \& Goldwurm 2003). The X-ray
source is transient in nature, and variability of the X-ray flux density on timescales of $\sim$ 2000 s
has been observed from the source along with the flaring events during its discovery (Rodriguez et
al. 2003). An EGRET gamma ray source  3EG 1631$-$4033 lies in the field direction at
a distance of 10$'$  with respect to the source, IGR J16316$-$4028 (Rodriguez {et al.} 2003) suggesting a
possible association of the two due to large position uncertainty of the gamma ray source.
During our observations with GMRT at 0.61 GHz and 1.28 GHz, no radio source was detected within the 
 X-ray position error box of 3$\sigma$ and upper limits of $\sim$ 0.8 mJy b$^{-1}$ and $\sim$ 6 mJy b$^{-1}$ 
respectively were measured. 
\\\\
\noindent{\bf  IGR J16318$-$4848 :} 
This transient source was discovered by {\emph{INTEGRAL}} in the Norma arm.
 The observed flux density in the 15 $--$ 40 keV band  was  50 $--$ 100 mCrab. The  X-ray flux data 
show significant temporal variations on the time scale of 1000 s. A bright 9th mag 
infrared source in the $J$, $H$, $K$ bands, coincident with the X-ray source position is clearly 
seen in the  2MASS and Mid course Space Experiment (MSX) data. The source is highly absorbed 
at 8 micron flux density and Galactic in nature (Courvoisier {et al.} 2003). 
The best fit position (J2000) determined from the infrared and the XMM measurement 
corresponds to  RA: 16h 31m 48.3s, DEC: $-$48d 49$'$ 01$'$$'$ (Schartel {et al.} 2003, 
de Plaa {et al.} 2003, Revnivtsev {et al.} 2003). The infrared and optical observations of 
the counterpart have revealed the high mass X-ray binary nature of the source, there being a compact, 
companion around a supergiant primary B[e] star (Filliatre {et al.} 2004). Strong photo absorption 
and the fluorescent emission line of neutral iron (6.4 keV) are seen in the XMM data thereby 
suggesting accretion due to a stellar wind (Matt {et al.} 2003). A search for the radio 
emission from the source at 4.8 and 8.6 GHz was made by Walter {et al.} 2003 with ATCA, 
but no significant radio flux density was detected from the source with a 1$\sigma$ upper 
limit of 0.1 mJy. We made repeated observations at 0.61 GHz using GMRT at various 
epochs during 2003 and 2004 observing cycles. No positive detection was made in any of these 
observations. An upper limit of $\sim$ 7 mJy for the radio emission within the 3$\sigma$ 
position uncertainty circle is derived. The typical rms noise in the field is measured to be 
$\sim$ 2.2 mJy. A non detection at radio frequencies may be due to the highly absorbed 
nature of the source due to its location in the Galactic plane. The present observation 
may also correspond to high soft state of the source during which the radio emission is believed 
to be quenched. There is also a possibility of non existence of the radio jet in the source, as 
in the case of pulsars.
\\\\
\noindent{\bf   IGR J16320$-$4751 :}
The variable hard X-ray source, IGR J16320$-$4751, was discovered during Feb 2003 
(Tomsick {et al.} 2003). The hard X-ray source position is consistent with that of 
the source AX J1631.9$-$4752. The ASCA/GIS observations also suggest a variable 
nature of the source (Sugizaki {et al.} 2001a) and its energy spectrum fits an 
absorbed power law and so it is believed to be a high-mass X-ray binary system. The variations 
of the spectral index in different observations by XMM-Newton, 
ASCA and BeppoSAX suggest that the source undergoes a spectral transition similar to 
those in the other XRBs. The EGRET source 3EG J1639$-$4702 (Romero {et al.} 1999) 
could be associated with the X-ray source considering the large position uncertainties.
The infrared images from the 2MASS database and the optical observations indicate 
a bright and massive companion for the X-ray source (Rodriguez {et al.} 2003).

GMRT observation of the field did not show any significant radio source within the 
3$\sigma$ position uncertainty limit during different observations at various frequencies. 
A 3$\sigma$ upper limit for the radio flux density was $\sim$ 3 mJy at 1.28 GHz with 
a rms noise of $\sim$ 0.83 mJy b$^{-1}$, while the 3$\sigma$ upper limit at 0.61 GHz correspond 
to 4 mJy with rms noise typically $\sim$ 1.04 mJy b$^{-1}$. No NVSS data is available for 
comparison of the field and also there is no RXTE/ASM data to determine the X-ray state 
of the source during our observation. The non detection of the source at low frequency 
clearly points out either an absorption of the jet in the ambient medium or a lack of a radio jet.
\\\\
\noindent{\bf IGR J16418$-$4532 :}
This hard X-ray source was discovered in the IBIS/ISGRI data at 
the flux density level of $\sim$ 3 mCrab in the 20 $--$ 40 keV bands (Tomsick {et al.} 2004). 
The X-ray source is located in the Norma arm region. The GMRT observations at 0.61 GHz were made 
in Feb 2004, to search for the radio counterpart 
of the X-ray source. No significant radio flux density was detected within the 
{\emph{INTEGRAL}} position uncertainty limit of 2$'$. The 3$\sigma$ upper limit is 1 mJy at 
1.28 GHz with a rms noise of $\sim$ 0.33 mJy b$^{-1}$. No RXTE/ASM X-ray light curve or the  
NVSS image is available. 
Thus suggesting a non existence of the radio emission from the X-ray source during our observations. 
\\\\
\noindent{\bf IGR J17391$-$3021 :}
Hard X-ray emission in the 18 $--$ 50 keV bands from the transient HMXB source, IGR J17391$-$3021,
was discovered in Aug 2003 at a flux density level of $\sim$ 70 mCrab
(Sunyaev {et al.} 2003a). The position coordinates of the source overlap with the X-ray
source XTE J1739$-$302 with J2000 position of, RA: 17h 39m 11.1s and DEC: $-$30d 20$'$ 37$'$$'$.
Using  CHANDRA observations, the  position has been refined to 1$'$$'$.  Within the
improved position, an optical counterpart of the source is identified with the
object 0525 28760590 in the USNO A2 catalog. It is also a 2MASS source and is a
highly reddened O star (Smith {et al.} 2003). A radio source was detected in a 
VLA observations at 4.9 GHz at a flux density of 0.9$\pm$0.2 mJy, and a J2000 position of, 
RA: 17h 39m 01.52s and DEC: $-$30d 19$'$ 34.9$'$$'$ within 3$'$ error circle {\emph{INTEGRAL}} 
position but well outside the CHANDRA error circle (Rupen {et al.} 2003b). The EGRET source 
3EG J1736$-$2908 is located near the X-ray source.

The source was observed with GMRT  at 1.28 GHz and no radio emission was
detected from the field within 3$\sigma$ position uncertainty of 2$'$$'$, as
determined from CHANDRA data. The 3$\sigma$ upper limit for the radio flux density
from GMRT data is $\leq$ 1 mJy and the rms noise was 0.2 mJy b$^{-1}$. No significant
radio flux from the X-ray source position was detected in the analysis of the
NVSS data and the 3$\sigma$ upper limit for the radio flux density is derived
as 2.5 mJy.

Nevertheless, during our observations, we did detect a radio source coincident in position
with the NVSS source with radio flux density of 8.9 mJy and rms noise of
0.19 mJy b$^{-1}$ at J2000 position of, RA: 17h 39m 06s and DEC: $-$30d 18$'$ 12$'$$'$ which
lies 2.4$'$ away from the X-ray source position. Thus we confirm from the above data
that the radio source detected by VLA in the field of the X-ray source is not associated
with the Galactic X-ray source.
\\\\
\noindent{\bf  IGR J17544$-$2619 :}
This transient source was discovered with {\emph{INTEGRAL}} at a flux
density of $\sim$ 160 mCrab in the 18 $--$ 25 keV bands (Sunyaev {et al.} 2003b). The X-ray
variability with time scales of 1.5 $--$ 2 hrs seen in the data is ascribed to the binary
nature of the X-ray source and is associated with the accretion-emission processes
(Rodriguez {et al.} 2003). A ROSAT source IRXS J175428.3$-$262035 with the
count rate of 0.08$\pm$0.02 counts s$^{-1}$ (0.1 $--$ 2.4 keV) lies within the field of
IGR J17544$-$2619. Optical and infrared parameters observed in the 2MASS survey
give the magnitude $J = 8.79\pm0.02$, $H = 8.31\pm0.03$, $K = 8.02\pm0.03$, thus
suggesting the companion star to be an early O-type star.\\
The GMRT image of the field of source IGR J17544$-$2619 at 0.61 GHz showed no
significant radio flux density within the 3$\sigma$ position uncertainty of the
X-ray source. A 3$\sigma$ upper limit for the radio emission from the GMRT data is
$<$7 mJy with the rms noise of $\sim$ 2.2 mJy b$^{-1}$. No radio emission from the source
region is seen even in the NVSS image. A 3$\sigma$ upper limit from the NVSS data
is estimated to be $\sim$ 3 mJy. The RXTE/ASM hardness ratio data for the source
suggests that it was in the high soft state during our observations. A non detection
of radio emission within the 4$'$$'$ (1.6$\sigma$) (Gonzalez-Riestra {et al.} 2004) position
limit from XMM-Newton observations, during a  high soft X-ray state of the source
suggests that there is no associated radio source for the X-ray source; however,
the radio emission, if present, may be synchrotron self-absorbed at low frequency or highly variable,
similar to some of the HMXBs below 1 GHz.
\\\\
\noindent{\bf  IGR J17597$-$2201 :}
 The X-ray source is a weak transient source with a flux density level 
of 5 mCrab in 15 $--$ 40 keV energy band (Lutovinov {et al.} 2003). 
The RXTE observation shows variability of the source.
The PCA spectra is, however, consistent with a power law with an additional 
soft X-ray excess. The measured spectral indices in the hard X-ray band 
range between 1.8 and 2.9. An Fe K emission line at 6.5 keV has been detected 
(Markwardt {et al.} 2003) in the X-rays. These observations 
suggest that IGR J17597$-$2201 is a binary system containing a neutron 
star as the compact companion. The GMRT observations on 16th Sep 2003 at 
0.61 GHz were taken immediately after the discovery of the source with {\emph{INTEGRAL}}.
The radio image of the field showed no significant radio source within the 
3$\sigma$ position uncertainty circle of the X-ray source position. A 3$\sigma$
upper limit for the radio flux density is derived as $<$7 mJy with the rms 
noise of $\sim$ 2.14 mJy b$^{-1}$. The analysis of the NVSS data also did not 
yield any radio flux density at 1.4 GHz and a 3$\sigma$ upper limit is 
obtained as 3.5 mJy. An upper limit for the radio emission from the source 
clearly implies that if the prompt radio emission was indeed present, 
then either it was highly absorbed at low frequency or was variable.
\\\\
\noindent{\bf  IGR J18325$-$0756 :}
This is a transient, flat spectrum X-ray source discovered with 
IBIS at a flux density level of $\sim$ 10 mCrab in the 15 $--$ 40 keV band and $\sim$ 5 mCrab 
in the 40 $--$ 100 keV band. The source is variable in X-ray regime (Lutovinov {et al.} 2003). 
We observed the source with GMRT at 0.61 GHz during AO3 to look for any 
residual low frequency flux density from the active phase. No significant radio source was 
detected within the 3$\sigma$ {\emph{INTEGRAL}} position error circle. The GMRT 3$\sigma$
upper limit for the radio flux density is $\leq$ 3 mJy. Thus our conclusion for non detection 
of radio emission is similar to IGR J17597$-$2201.
\\\\
\noindent{\bf IGR J18483$-$0311 :}
This new transient source was discovered in Apr 2003 with a 
X-ray flux density of $\sim$ 10 mCrab in the 15 $--$ 40 keV band and a flaring behavior 
in the X-ray light curve was seen from the source (Chernyakova {et al.} 2003).
The radio image of the field of IGR J18483$-$0311 with GMRT at 1.28 GHz showed
no significant radio source within the 3$\sigma$ upper limit of 4 mJy either at 
0.61 GHz or 1.28 GHz. No significant source was detected even in the NVSS data. 
The 3$\sigma$ upper limit at 1.2 GHz was 1 mJy. A non detection of the radio emission 
supports our previous conclusion.
\\\\
\noindent{\bf  IGR J19140$+$0951 :}
The X-ray source was discovered with IBIS/ISGRI at a flux density 
of $\sim$ 50 $--$ 100 mCrab between 15 and 40 keV (Hannikainen {et al.} 2003). The RXTE 
light curve of the source shows variations on short time scale. No QPO was seen in the 
power density spectrum of the source in the PCA/RXTE data, the X-ray spectrum is; 
however, well fitted by power-law $+$ thermal component (Cabanac {et al.} 2004a).
A broad ionized iron line is also seen in the spectral data.
The source shows strong spectral variability more typical of X-ray binaries. 
A comparison of the ISGRI/JEMX and RXTE/PCA spectra reveals a state 
transition in the source from `steep power-law' to `low hard'
(Schultz {et al.} 2004). Such a transition in spectral state is a characteristic of
a BH LMXB (McClintock {et al.} 2003) thus suggesting the possibility of
IGR J19140$+$0951 being a BH LMXB, though the absence of QPO 
does not fit this hypothesis. However; Rodriguez (private communication) suggests the HMXB nature
of the source, with a neutron star as the central engine based on available RXTE and {\emph{INTEGRAL}} data.
We performed a series of ten radio observations  with GMRT for the source at 0.61 and 1.28 GHz
during the 2003 and 2004 observing seasons. No positive radio detection  within the 1.6$\sigma$
position error circle of 1.3$'$ around the new refined position coordinate (Cabanac {et al.} 2004b)
of the source was found during our observations.
The upper limit for the GMRT radio flux density towards the source direction is derived as $\leq 5.5$ mJy.
No radio source corresponding to the X-ray source position was seen even in the NVSS data at 1.4 GHz. 
The absence of radio emission from the source as seen in other microquasars, viz XTE J1748$-$288, 
GRS 1758$-$258 below 1.4 GHz is also seen in IGR J19140$+$0951. Thus we confirm from our observations 
that the X-ray source has no radio counterpart, or if present, it is highly absorbed below 1.4 GHz.

\section{Discussion} 
Among the seventeen sources for which GMRT data is  presented in the previous section, 
radio counterparts are detected for seven objects that lie within the 3$\sigma$ position 
error circle of the {\emph{INTEGRAL}} source positions. Among these seven radio sources, 
the morphology for IGR J16479$-$4514 resembles a HII region while that of IGR J18539$+$0727 and 
IGR J21247$+$5058 indicates a radio galaxy. In the case of IGR J15479$-$4529, GMRT data do not fit a point source 
model, thereby suggesting an extended morphology more likely to be an extragalactic object. 
The sources IGR J06074$+$2205 and IGR J18027$-$1455 are unresolved source with extragalactic 
properties and are probably AGNs. The {\emph{INTEGRAL}} source, IGR J17091$-$3624 for which 
coincident radio emission is measured from the GMRT data, qualify as a  binary source and a 
possible microquasar candidate.

The sources IGR J16318$-$4848, IGR J16320$-$4751, IGR J17544$-$2619, IGR J17597$-$2201 and 
IGR J19140$+$0951 are suggested as X-ray binaries from the {\emph{INTEGRAL}} studies; however, 
their radio counterparts are not yet detected, suggesting that either they are synchrotron 
self absorbed at low frequencies or have no current radio emission. The sources IGR J16316$-$4028, 
IGR J16418$-$4532, IGR J17391$-$3021, IGR J18325$-$0756 and IGR J18483$-$0311 do not have low 
frequency radio counterparts and were not identified based on our survey. This points to either 
a variable nature of the radio emission or a persistent emission at much lower flux density 
than the GMRT threshold. As seen in Table 2, the position off-set of the possible radio 
counterparts with respect to the X-ray source position are of the order of few arcminutes and 
well within the 3$\sigma$ error circle of the X-ray position. 
Hence such possible identifications are a reason to conduct even deeper optical and infrared 
follow-up observations.

A large number of accreting binary systems show a transient behavior in the X-ray band. 
The radio emission from X-ray binaries can be classified into two broad classes  
(1) X-ray sources with  persistent radio emission are generally binary systems 
containing a BH or a NS with low magnetic field; 
(2) the sources with a highly variable and bright radio emission, are 
associated with jet-like outflows from their compact objects.

On the basis of X-ray spectral characteristics, their transient nature and their 
discovery in the high energy band, a majority of the {\emph{INTEGRAL}} sources are believed to be 
Galactic HMXB. It is, therefore likely that during the transient outbursts, 
most of these sources will produce relativistic outflows, thereby giving rise to strong radio emission. 
Therefore, non-detection of radio flux density from a number of {\emph{INTEGRAL}} 
sources may be due to the different X-ray state of the source at the time of our observations. 
Thus, near simultaneous observations in different wave bands are necessary to reveal the 
true nature of most of the {\emph{INTEGRAL}} high energy sources.

\begin{acknowledgement}
We  acknowledge the excellent support provided by the GMRT staff during our observations. 
We are also thankful to our colleagues at TIFR and NCRA  for helpful discussions. 
This research had made extensive use of the NASA/IPAC extragalactic sources data base NED, 
Data from the NASA/NVSS survey and the public archives from the RXTE and {\emph{INTEGRAL}} Satellites.
MP is grateful to Director NCRA  for providing the research facilities in Pune and is also 
thankful Prof. V. H. Kulkarni of Mumbai University  for his constant support and encouragement. 
This research is supported by ISRO, grant No. DOS. 9/2/109/2003-II under its RESPOND program. 
MP also thanks Prof. Kulkarni V., Witta P., Green D., Rodriguez J., and Ribo M. for
useful discussions. 
\end{acknowledgement}

\cite{whatever}

\bigskip
\vspace*{2.6cm}
\newpage
\begin{table*}
\begin{center}
\caption{ Possible Radio counterparts of target INTEGRAL sources observed with GMRT}
\begin{tabular}{lllllllllll}
\hline
\hline
Source         &Date    &$\nu$&$S\sb{\nu}$   &$\sigma$       &Radio    &Pos. Off.&Radio    &$S\sb{\nu}$\\
               &dd/mm/yy&     &              &(mJy  b$^{-1}$)&Pos. GMRT&w.r.t    &Struc.   &(Total)    \\
               &        &(GHz)&(mJy)         &               &RA \& DEC&Integral &         &NVSS       \\
               &        &     &              &               &         &Pos.     &GMRT     &1.4 GHz    \\
               &        &     &              &               &         &         &         &(mJy)    \\
\hline
IGR J06074$+$2205&01/02/04&1.28&36.00$\pm$1.12 &0.64         &06h 07m 18.45s$\pm$0.005       &1.29$'$&Point   &36\\
             &        &     &              &                 &22d 04$'$ 52.49$'$$'$$\pm$0.03 &    &           & \\
               &09/04/04&0.61 &78.65$\pm$1.29 &1.33          &                               &    &           & \\
               &12/04/04&0.61 &83.49$\pm$1.30 &0.65          &                               &    &           & \\
               &02/05/04&0.61 &79.86$\pm$1.42 &1.57          &                               &    &           & \\
IGR J15479$-$4529&02/02/04&1.28 &22.2 &0.5                   &15h 47m 44.26s$\pm$0.105       &3.98$'$&Extended&N/A\\
           &        &     &              &                   &$-$45d 32$'$ 35.22$'$$'$$\pm$1.76&  &           & \\
IGR J16316$-$4028&02/02/04&1.28 &$\le$0.8   &0.19            &                               &    &           & \\
               &        &     &              &               &              &    &                            & \\
               &09/04/04&0.61 &$\le$6.00 &1.35               &              &    &                            & \\
               &12/04/04&0.61 &$\le$4.80 &1.18               &              &    &                            & \\
IGR J16318$-$4848&04/08/03&0.61 &$\le$6.50     &2.13         &              &    &                            &N/A\\
               &15/09/03&0.61 &$\le$6.60     &2.20           &              &    &                            &  \\
               &09/04/04&0.61 &$\le$7.00     &2.41           &              &    &                            &  \\
IGR J16320$-$4751&29/07/03&1.28 &$\le$3.00     &0.83         &              &    &                            &N/A\\
               &04/08/03&0.61 &$\le$4.00     &1.23           &              &    &                            &  \\
               &15/09/03&0.61 &$\le$4.00     &1.04           &              &    &                            &  \\
IGR J16418$-$4532&02/02/04&1.28 &$\le$1.00     &0.33         &              &    &                            &N/A\\
IGR J16479$-$4514&02/02/04&1.28 &2445.45       &1.24         &16h 47m 47.75s$\pm$0.153&3.30$'$       &Extended&N/A\\
            &        &     &              &                  &$-$45d 17$'$ 07$'$$'$$\pm$1.97&        &        &  \\
IGR J17091$-$3624&16/09/03&0.61 &2.50$\pm$0.73 &0.60         &17h 09m 02s$\pm$0.191       &0.99$'$   &Point   &1.5\\
           &        &     &              &                   &$-$36d 23$'$ 33$'$$'$$\pm$2.84&        &        &  \\
               &02/02/04&1.28 &1.10$\pm$0.16 &0.19           &                              &        &        &  \\
IGR J17391$-$3021&02/02/04&1.28 &$\le$1.04   &0.19           &                              &        &    &$\le$2.5\\
IGR J17544$-$2619&12/04/04&0.61 &$\le$7.35     &2.46         &                              &        &    &$\le$3\\
IGR J17597$-$2201&16/09/03&0.61 &$\le$7.90     &2.42         &                              &        &    &$\le$3.5\\
IGR J18027$-$1455&02/05/04&0.61 &10.72$\pm$2.25&2.32         &18h 02m 42.75s$\pm$0.134      &4.18$'$ &Point   &12.3\\
           &        &     &              &                   &$-$14d 50$'$ 49.21$'$$'$$\pm$3.93&     &        & \\
IGR J18325$-$0756&16/09/03&0.61 &$\le$3.00     &0.75         &                              &        &    &$\le$3.5\\
IGR J18483$-$0311&16/09/03&0.61 &$\le$4.24     &1.27         &                              &        &        &  \\
               &02/02/04&1.28 &$\le$1.00     &0.15           &                              &        &    &$\le$1.6\\
IGR J18539$+$0727&02/02/04&1.28 &5.25$\pm$0.25$\pm$0.14 &0.16&18h 53m 52.54s$\pm$0.028      &3.64$'$ &Double  & \\
               &        &     &              &               &$+$07d 30$'$ 39.07$'$$'$$\pm$0.43&     &Point   &  \\
               &        &     & 5.85$\pm$0.50&               &18h 53m 52.19s$\pm$0.023      &        &        &6 \\
               &        &     &              &               &$+$07d 30$'$ 37.31$'$$'$$\pm$0.32&     &        &  \\
IGR J19140$+$0951&21/06/03&0.61 &$\le$5.50  &1.28            &                                 &     &    &$\le$1.6\\
             &29/07/03&1.28 &$\le$0.80     &0.15             &                                 &     &        &\\
             &04/08/03&0.61 &$\le$5.50     &1.23             &                                 &     &        &\\
             &09/09/03&0.61 &$\le$5.00     &1.14             &                                 &     &     &  &\\
             &15/09/03&0.61 &$\le$5.70     &1.90             &                                 &     &        &\\
             &02/02/04&1.28 &$\le$0.50     &0.15             &                                 &     &     &  &\\
             &07/04/04&0.61 &$\le$5.00     &1.43             &                                 &     &     &  &\\
             &12/04/04&0.61 &$\le$5.00     &1.28             &                                 &     &     &  &\\
             &02/05/04&0.61 &$\le$5.00     &1.64             &                                 &     &     &  &\\
             &03/05/04&0.61 &$\le$5.00     &1.20             &                                 &     &     &  &\\
IGR J21247$+$5058&07/04/04&0.61 &184.81$\pm$1.59&1.25        &21h 24m 39.63s$\pm$0.004      &0.57$'$ &Extended&237\\
              &        &     &               &               &50d 58$'$ 25.55$'$$'$$\pm$0.04   &     &        &\\
              &09/04/04&0.61 &178.41$\pm$1.38&1.61           &                                 &     &        &\\
              &12/04/04&0.61 &173.81$\pm$1.59&1.71           &                                 &     &        &\\
              &02/05/04&0.61 &177.71$\pm$1.17&1.52           &                                 &     &        &\\
\hline
\end{tabular}
\end{center}
\end{table*}

\begin{thebibliography}{}
\bibitem{whatever} Basko M., Sunyaev R., Titarchuk L., 1974, Astron. Astroph. 31, 249
\bibitem{whatever} Belloni T., Mendez M., King A., et al., 1997, ApJ, 488, L109
\bibitem{whatever} Bird A., Barlow E., Bassani L., et al., 2004, ApJ, 607, L33
\bibitem{whatever} Barth A., Luis H., Filippenko A., et al., 2001, ApJ., 546, 205-209 
\bibitem{whatever} C. O' Dea, et al., 1998, ApJ, 488, L109
\bibitem{whatever} Cabanac C., Rodriguez J., Hannikainen D., et al., 2004a, ATEL 272
\bibitem{whatever} Cabanac C., Rodriguez J., Petrucci P., et al., 2004b, RMxAC, 20, 209
\bibitem{whatever} Ceballos M. and Barcons X., 2004, Multiwavelength AGN surveys conference proceedings, Cozumel, 17
\bibitem{whatever} Chenevez  J., Budtz-Jorgensen C., Lund N., et al., 2004, ATEL 223
\bibitem{whatever} Chernyakova M., Lutovinov A., et al., 2003, ATEL 157
\bibitem{whatever} Combi, Ribo, Mirabel, et al., 2004, ATEL 246
\bibitem{whatever} Condon J. Mirabel, 1999, Proc. Natl. Acad. Sci., Vol. 96, pp. 4756-4758
\bibitem{whatever} Cordier B., Goldwurm A., Laurent P., et al., 1991, Adv.Sp.Res. 11, 169
\bibitem{whatever} Courvoisier T., Walter R., Rodriguez J., et al., 2003, IAUC 8063
\bibitem{whatever} de Plaa J., Hartog den P., Kaastra J., et al., 2003, ATEL 119
\bibitem{whatever} Fenimore E., Cannon M., 1978, Appl. opt. 17, 337
\bibitem{whatever} Filliatre \& Chaty, 2004, ApJ, 616, 496
\bibitem{whatever} G. Matt \& Guainazzi, et al., 2003, MNRAS, 341, L 13  
\bibitem{whatever} Gonzalez-Riestra E., Oosterbroek T., Kuulkers E., et al., 2004, A\&A, 420, 589
\bibitem{whatever} Hannikainen D., Rodriguez J., Pottschmidt K., et al., 2003, IAUC 8088
\bibitem{whatever} Haberl F., Motch C., Zickgraf F., 2002, A\&A., 387, 201H
\bibitem{whatever} Ishwara-Chandra, A. P.Rao, Pandey M., et al., 2005, ChJAA, 5, 87-92 
\bibitem{whatever} Kuulkers E., Lutovinov A., Parmar A., et al., 2003, ATEL 149
\bibitem{whatever} Lebrun  F., Leray J., Lavocat P., et al., 2003, A\&A, v.411, pL141-148
\bibitem{whatever} Lutovinov A., Rodriguez J., et al., 2003, ATEL 151
\bibitem{whatever} Lutovinov A., Shaw S., et al., 2003, ATEL 154
\bibitem{whatever} Lutovinov A., Walter R., et al., 2003, ATEL 155
\bibitem{whatever} Lutovinov A., Rodriguez J., Produit N., et al., 2003, ATEL 151
\bibitem{whatever} Malizia A., Bassani L., et al., 2004, ATEL 227
\bibitem{whatever} Markoff S. \& Nowak M. 2004, ApJ, 609, 972
\bibitem{whatever} Markwardt, Swank J., et al., 2003, ATEL 156
\bibitem{whatever} Masetti N., Palazzi E., Bassani L., et al., 2004, A\&A, 426L, 41
\bibitem{whatever} McClintock J., Remillard R. 2003, `Compact Stellar X-ray sources, eds, Cam. Univ. Press'
\bibitem{whatever} Mirabel F., Gerard E., Rodr\'{\i}guez L. 1994, IAU Circ., 5958
\bibitem{whatever} Mirabel F., Harmon B., Deal K., et al., 1997, ApJ, 477, L85
\bibitem{whatever} Mirabel F., Marti J., Chaty S., et al., 1998, New Aston. Rev., 42, 9-10, 621-624
\bibitem{whatever} Mirabel F. and Rodr\'{\i}guez L. 1999, ARA\&A, 37, 409
\bibitem{whatever} Molkov S., Mowlavi N., Goldwurm A., et al., 2003, ATEL 176
\bibitem{whatever} Pandey M., Durouchoux Ph., Manchanda R., et al., 2004, 5th INTEGRAL workshop 
proceedings held in Munich, ESA, SP-552, 695
\bibitem{whatever} Pooley G. 2004, ATEL 226
\bibitem{whatever} Revnivtsev M., Astron. Astroph. 2003a, A\&A, 411, 329
\bibitem{whatever} Revnivtsev M., Lutovinov A. 2003b, Astron. Letters, Vol. 29, No. 11, pp.719-723
\bibitem{whatever} Rodriguez J., Foschini L., Goldwurm A., et al., 2004, ATEL 253
\bibitem{whatever} Rodriguez J., Goldwurm A. 2003, ATEL 201
\bibitem{whatever} Rodriguez J., Tomsick J. 2003, Foschini L., IAUC 8096
\bibitem{whatever} Rodriguez J., 2003, ATEL 194
\bibitem{whatever} Romero A., Gaztanaga E., Barriga., et al., 1999, A\&A, 348
\bibitem{whatever} Ribo M., Combi J., 2003, ATEL 235
\bibitem{whatever} Rupen M., Mioduszewski A., Dhawan V. 2003a, ATEL 152
\bibitem{whatever} Rupen M., Mioduszewski A., Dhawan V. 2003b, ATEL 184
\bibitem{whatever} Schartel N., Ehle M., Breitfellner M., et al., 2003, IAUC 8072
\bibitem{whatever} Schultz J., Hannikainen D., Vilu O., et al., 2004, A\&A, 423L, 17
\bibitem{whatever} Smith D., Heindl W., Swank J., et al., 2003, ATEL 182
\bibitem{whatever} Sugizaki M., Matsuzaki K., Kaneda H., et al. 2001(a), AIPC, 599, 959
\bibitem{whatever} Sugizaki M., Matsuzaki K., Kaneda H., et al. 2001(b), Ap.J. Suppl. 134, 77
\bibitem{whatever} Sunyaev R., Lotovinov A., et al. 2003a, ATEL 181
\bibitem{whatever} Sunyaev R., Grebenev S., Lutovinov A., et al., 2003b, ATEL 190
\bibitem{whatever} Sunyaev R., Titarchuk, et al., 1980
\bibitem{whatever} Swarup G. 1991, Radio Interferometry techn.:Proceedings of 131st IAU Coll.,Astr. 
Scty. Pac., p.376-380
\bibitem{whatever} Tomsick J., Lingenfelter R., et al., 2003, IAUC 8076
\bibitem{whatever} Tomsick J., Lingenfelter R., Corbel S., et al., 2004, ATEL 224
\bibitem{whatever} Torres D., Romero G., Dame T., et al., 2003, Phys. Rep., 382, 303
\bibitem{whatever} Ubertini P., Bazzano A., Cocchi M., et al., 1999, Astroph.Lett.Comm. 38, 301 
\bibitem{whatever} Van der Klis M., Wijnands R. 1999, Astrophys.J. 514, 939
\bibitem{whatever} Walter R., Bodaghee A., et al., 2004, ATEL 229
\bibitem{whatever} Wright \& Barlow, 1975
\bibitem{whatever} Zand J. J. M., Heise J., et al., 2003, ATEL 229
\end{thebibliography}
\end{document}